\documentclass[prd,onecolumn,nofootinbib,showpacs,showkeys]{revtex4}  

\usepackage{latexsym,graphicx,amssymb,amsmath,mathrsfs}  

\usepackage{setspace,bm}  
      
\newcommand{\be}{\begin{equation}}      

\newcommand{\ee}{\end{equation}}      

\newcommand{\bea}{\begin{eqnarray}}      

\newcommand{\eea}{\end{eqnarray}}

\newcommand{\rt}[1]{{}}

\makeatletter
\renewcommand\appendix{\par
\setcounter{section}{0}%
\setcounter{subsection}{0}%
\gdef\thesection{\appendixname\space\@Alph\c@section}}
\makeatother

\begin{document}           
{\allowdisplaybreaks   
  
\title{A renormalized large-$n$ solution \\    
of the $U(n)\times      
  U(n)$  linear sigma model\\ in the broken symmetry phase}        
  
\author{G. Fej\H{o}s}
\email{geg@ludens.elte.hu}
\affiliation{Department of Atomic Physics, E{\"o}tv{\"o}s University,
H-1117 Budapest, Hungary}
  
\author{A. Patk{\'o}s}
\email{patkos@galaxy.elte.hu}
\affiliation{Department of Atomic Physics, E{\"o}tv{\"o}s University,
H-1117 Budapest, Hungary}
\affiliation{Research Group for Statistical and Biological Physics 
of the Hungarian Academy of Sciences, H-1117 Budapest, Hungary}
    
\begin{abstract}        
{Dyson-Schwinger equations for the $U(n)\times U(n)$ symmetric        
matrix sigma model reformulated with two auxiliary fields 
 in a background breaking the symmetry to    
$U(n)$ are studied in the so-called bare vertex approximation. A large $n$ solution 
is constructed under the supplementary assumption so that  
the scalar components are much heavier than the pseudoscalars.         
The renormalizability of the solution is investigated by explicit construction of the counterterms.     
}        
\end{abstract}       
\maketitle         
  
\section{Introduction}

Universal features of finite temperature and  
finite density variation of the QCD ground state realizing
 the $SU(3)_R\times SU(3)_L$ approximate  
chiral symmetry were understood with the help of  
the corresponding meson model \cite{pisarski84}. There were also numerous  
attempts to treat quantitatively the finite temperature    
restoration of chiral symmetry of the  
strongly interacting matter in the framework of this model.   
A common goal of these investigations is the description of the quark mass dependence of the  
nature of the finite temperature symmetry restoration   
\cite{lenaghan00,roder03,herpay05}.   
Another central issue is the nature of the phase transition, when the   
baryonic density is varied. For its investigation one couples  
constituent quarks carrying baryon number to the meson model  
\cite{bilic98,kovacs09,schaefer09,schaefer10}.  
  
Optimistically one could say, that the location of the characteristic
points of the QCD phase diagram determined with different variants of the  
model  and in  different approximations   
do agree with each other and with the results of   
the lattice field theoretical simulations  
within a factor of 2. In particular, improved agreement with lattice
determinations of the QCD phase diagram were reported, when the
Polyakov loop degree of freedom is coupled to the quark-meson
model \cite{kahana08,mao10,gupta10}. But the effective models are all strongly  
coupled, therefore one usually experiences large variations  
in their predictions, when the simplest mean-field treatments are
improved by taking into account quantum fluctuations of the mesons and
the constituent quarks \cite{marko10,herbst10}. This circumstance
limits the competitivity of their predictions.  
  
Some time ago we initiated the application of the expansion in  
the number of quark flavors for the description of the chiral  
symmetry restoration in the two-flavor model \cite{patkos02,jakovac04}. The approach was suggested by the $SU(2)\times SU(2)\sim O(4)$ isomorphy and the  fortunate fact that the $N=\infty$ solution of the 
 $O(N)$ model is the textbook example of the application  
of the $1/N$ expansion \cite{zinn-justin02}. Recently, next-to-leading  
order (NLO) results were also presented for the pressure of the  
relativistic $O(N)$ model and applied to the physical ($N=4$)  
sigma-pion gas \cite{andersen04,andersen08}. In this context also the  
renormalizability of the NLO approximation was fully clarified and  
shown to be valid in various formulations of the model (with and  
without auxiliary field) \cite{fejos09}.  
  
An analogous development for the three-flavor meson (and quark-meson)  
model is hindered, because even the leading order solution of  
the large-$n$ limit of the $U(n)\times U(n)$ symmetric linear  
sigma-model is unknown. Progress would represent interest also for
the Higgs sector of technicolor models of
electroweak symmetry breaking \cite{appelquist95,kikukawa07}.  
To our best knowledge, no published attempts exist, which would go
beyond the partial large-$n$ treatments of the $O(2n^2)$-symmetric
nonlinearities. The results obtained with such an approach are
questionable in light of the rather different finite temperature 
renormalization group behavior of the $O(2n^2)$ and $U(n)\times U(n)$
symmetric models for $n\geq 3$ \cite{pisarski84,paterson81}.  
  
The goal of the present paper is to describe an  
approximate leading order ($n=\infty$)   
solution. Although it takes into account only two-loop contributions to the 2PI (two-particle irreducible) effective action, it definitely goes beyond the $O(2n^2)$ symmetric solution. It exploits  
in addition to the large $n$ expansion an assumption concerning the  
mass spectra of the model. The region in the coupling space,
where this assumption is valid can be estimated by determining the  
spectra from the approximate solution self-consistently. 
In this paper we present the  
construction of the renormalized version of this solution in some detail, and provide also an
illustrative investigation of its range of validity. A deeper analysis of the mass
spectra and the finite temperature features are left to a forthcoming publication. 
  
The idea to impose an extra assumption on the mass spectra
stems from the common practice of  
dealing with the 4 scalars and 4 pseudoscalars defining the $SU(2)\times  
SU(2)$ symmetric meson model. One simply omits half of the fields,  
e.g. the 3 components of the  
scalar-isovector triplet and the pseudoscalar-isoscalar singlet with  
reference to their higher mass. One arrives this way
 at the Gell-Mann--L{\'e}vy  
linear sigma model. We shall assume an analogous feature to occur   
in the spectra of the leading order solution at large $n$. We
 do not attempt the anomalous realization of the $U_A(1)$ symmetry, therefore
 in this limiting case all pseudoscalars will have the same mass.  
Based on this assumption, in a first approximation to the $n=\infty$ 
solution we  
retain only the quantum fluctuations of the light pion fields.  
After finding the propagators of the scalar fields, one calculates corrections to the pion propagator
arising from the heavy  
scalar fluctuations. In principle, one can iterate this procedure  
until the solution of the large-$n$ bare vertex approximation (BVA) is reached.   
It will be demonstrated, that the pion fields still  
obey Goldstone's theorem after the scalar corrections are  
included. We shall explore the divergence structure of the equation of  
state, the pion self-energy and the saddle-point equations (SPEs),
and determine the counterterm pieces in the effective action necessary
for their renormalization.   
  
The paper is organized as follows. In Sec. II, the model is  
reformulated with two auxiliary fields. We elaborate also on
 the expected mixing structure of the solution. In Sec. III,  
the leading large-$n$ Dyson-Schwinger (DS) equations are presented in 
BVA and the corresponding 2PI 
effective potential is constructed. In Sec. IV, we present an
approximate solution, which is valid when the scalars are much heavier than the pseudoscalar fields of the model.
Here also a simplified exploration of the range of validity of this assumption is given.  
The renormalizability of the proposed solution is investigated in
Sec. V. In the divergence  
analysis we rely on a number of cutoff sensitive integrals collected in  
Appendix A. The paper ends with the conclusions (Sec. VI).

\section{Formulation of the model with auxiliary fields}          
The Lagrangian of the model in its usual form reads as        
\be        
L=\textrm {Tr}[\partial_\mu M\partial^\mu M^\dagger-m^2MM^\dagger]-        
\frac{g_1}{n^2}        
(\textrm {Tr}MM^\dagger)^2-\frac{g_2}{n}\textrm {Tr}(MM^\dagger)^2+        
\sqrt{2n^2}h s^0,        
\ee        
where the algebra valued complex field $M$ is parametrized with help        
of the generators of the $U(n)$ group:        
\be        
M=(s^a+i\pi^a)T^a,\qquad {\textrm        
  {Tr}}T^aT^b=\frac{1}{2}\delta^{ab},\qquad a=0,1,...,n^2-1.        
\ee        
In terms of the matrix elements, it can be written in the following form:       
\bea        
L&=&\frac{1}{2}[(\partial_\mu        
s^a)^2+(\partial_\mu\pi^a)^2-m^2((s^a)^2+(\pi^a)^2)]+\sqrt{2n^2}h s^0        
\nonumber\\        
&-&\frac{g_1}{4n^2}\Big((s^a)^2+(\pi^a)^2\Big)^2-\frac{g_2}{2n}(U^a)^2,        
\eea        
where the $U(n)$ vector $U^a$ is defined as        
\be        
U^a=\frac{1}{2}d^{abc}(s^bs^c+\pi^b\pi^c)-f^{abc}s^b\pi^c,       
\ee        
with $d^{abc}$ and $f^{abc}$ being the symmetric and antisymmetric
structure constants of the $U(n)$ group, respectively. Some useful
relations involving them can be found in Appendix B., which are
needed in order to obtain the subsequent equations.
Two auxiliary fields are introduced by adding the following 
constraints to the Lagrangian:        
\be        
\Delta        
L=-\frac{1}{2}\left(X-i\sqrt{\frac{g_1}{2n^2}}((s^a)^2+(\pi^a)^2)        
\right)^2-\frac{1}{2}\left(Y^a-i\sqrt{\frac{g_2}{n}}U^a\right)^2.        
\ee        
In the sum ${\cal L}\equiv L+\Delta L$ the $U(n)\times U(n)\rightarrow U(n)$ symmetry breaking pattern corresponds to the shifts    
\be        
s^a\rightarrow s^a+\sqrt{2n^2}v\delta^{a0},\qquad U^a\rightarrow        
U^a+2\sqrt{n}vs^a+n\sqrt{2n}\delta^{a0}v^2.        
\ee        
After the introduction of the auxiliary field variables and        
the shifts, the full Lagrangian has the following form:        
\bea        
{\cal L}&=&        
\frac{1}{2}\Big[(\partial_\mu s^a)^2+(\partial_\mu\pi^a)^2-m^2\Big(2n^2v^2
+  2\sqrt{2n^2}vs^0+(s^a)^2+(\pi^a)^2\Big)\Big]+\sqrt{2n^2}h (s^0+        
\sqrt{2n^2}v)\nonumber\\        
&-&\frac{1}{2}X^2-\frac{1}{2}(Y^a)^2+i\sqrt{\frac{g_1}{2n^2}}X\Big(2n^2v^2+        
2\sqrt{2n^2}vs^0+(s^a)^2+(\pi^a)^2\Big)\nonumber\\        
&+&i\sqrt{\frac{g_2}{n}}Y^a(U^a+2\sqrt{n}vs^a+\sqrt{2n^3}\delta^{a0}v^2).        
\label{Eq:shifted-aux-L}        
\eea          
      
Next, we shortly describe the assumed structure of the solution and introduce some notations.    
The classical constraint equations show that the background $v$        
induces nonzero $X,Y^0$ values and introduces mixing of the 
pair $s^0,X$ and also of $s^a,Y^a$ for every value of the index $a$. 
We construct correspondingly a quantum solution, where the saddle-point        
values of $X, Y^0$ are nonzero, $Y^a=0, a\neq 0$, and only        
the following 2-point functions do not vanish:        
\bea        
&\displaystyle        
[G_{s^0X},\quad G_{s^0Y^0},\quad G_{XY^0},\quad G_{s^0s^0},\quad        
G_{Y^0Y^0},\quad G_{XX}],\nonumber\\        
&        
\displaystyle        
[G_{s^us^v, u=v\neq 0},\quad G_{s^uY^v, u=v\neq 0},        
\quad G_{Y^uY^v, u=v\neq 0}], \qquad G_{\pi^u\pi^v, u=v}.        
\eea        
The sets in square brackets form mixing sets of fields: there is a 
3-dimensional mixing sector, and        
there are $n^2-1$ identical copies of 2-dimensional mixing two-point functions. The $\pi$ sector is diagonal.          
The equations presented in the next section show degeneracy of the        
2-point functions for $u=v\neq 0$, therefore it is convenient to introduce   
the following short-hand notations:        
\bea        
&        
\displaystyle        
G_{s^0s^0}\equiv G_{s^0}, \quad G_{XX}\equiv G_{X}, \quad G_{Y^0Y^0}\equiv G_{Y^0}, \quad G_{\pi^0\pi^0}\equiv G_{\pi^0},        
\nonumber\\        
&        
\displaystyle        
\quad G_{\pi^u\pi^u}\equiv G_{\pi},\quad G_{s^us^u}\equiv G_s,        
\quad G_{s^uY^u}\equiv G_{sY},\quad G_{Y^uY^u}\equiv G_Y,\qquad        
u\neq 0.        
\eea          
 
\section{Dyson-Schwinger equations and their truncation}            
\subsection{The equation of state and the saddle-point equations}        
      
Standard   rules \cite{rivers} of constructing the derivatives of the quantum effective        
action $\Gamma$ with respect to the fields leads to the following        
expressions:        
\bea        
\frac{\delta\Gamma}{\delta X}&=&-X+i\sqrt{\frac{g_1}{2n^2}}        
\Big(2n^2v^2+2\sqrt{2n^2}vs^0+(s^a)^2+G_{s^as^a}+(\pi^a)^2+G_{\pi^a\pi^a}\Big),        
\nonumber\\        
\frac{\delta\Gamma}{\delta Y^u}&=&-Y^u+i\sqrt{\frac{g_2}{n}}        
\Big(U^u+2\sqrt{n}vs^u+\sqrt{2n^3}\delta^{u0}v^2+\frac{1}{2}d^{ubc}        
(G_{s^bs^c}+G_{\pi^b\pi^c})-f^{ubc}G_{s^b\pi^c}\Big),        
\nonumber\\        
\frac{\delta\Gamma}{\delta s^u}&=&-(\square        
+m^2)s^u-\Big(m^2\sqrt{2n^2}-2i        
\sqrt{g_1}X\Big)v\delta^{u0}+\sqrt{2n^2}h \delta^{u0}+i\frac{\sqrt{2g_1}}{n}(Xs^u+    G_{Xs^u})        
\nonumber\\        
&+&i\sqrt{\frac{g_2}{n}}\Big[d^{abu}(s^bY^a+G_{s^bY^a})+2\sqrt{n}vY^u-f^{auc}        
(\pi^cY^a+G_{\pi^cY^a})\Big],\nonumber\\        
\frac{\delta\Gamma}{\delta \pi^u}&=&-(\square +m^2)\pi^u+        
i\frac{\sqrt{2g_1}}{n}(X\pi^u+G_{X\pi^u})        
+i\sqrt{\frac{g_2}{n}}\Big[d^{abu}(\pi^bY^a+G_{\pi^bY^a})-f^{abu}        
(s^bY^a+G_{s^bY^a})\Big].      
\label{Eq:1deriv}  
\eea        
Taking further functional derivatives of these expressions, one arrives        
at the equations of the 2-point functions (see next subsection).        
Also when one equates the derivatives of $\Gamma$ to zero,        
the equations determining the physical value of the background $v$        
and the        
auxiliary fields $X,Y^0$ are obtained (we do not distinguish the notations of the solutions        
of these equations from the variables taking arbitrary values). Using the  
Fourier representation of the propagators, from ({\ref{Eq:1deriv}}) one gets the two SPEs and
the equation of state (EoS) (setting $u=0$) at the leading order of the large $n$ expansion as  
\bea       
0&=&-X+in\sqrt{2g_1}v^2+i\sqrt{\frac{g_1}{2}}n\left(\int_k G_s (k)+\int_k G_{\pi}(k)\right),        
\nonumber\\        
0&=&-Y^0+in\sqrt{2g_2}v^2+i\sqrt{\frac{g_2}{2}}n\left(\int_k G_s (k)+\int_k G_{\pi}(k)\right),     
\nonumber\\
0&=&-\sqrt{2}nv\Big(m^2-i\frac{\sqrt{2g_1}}{n}X-i\frac{\sqrt{2g_2}}{n}Y^0\Big)+\sqrt{2}nh+i\sqrt{2g_2}n\int_k G_{sY}(k),      
\label{Eq:1pointeq}         
\eea    
respectively. We note, that the equations for $\pi^a$        
and $s^u$ imply for $u\neq 0$ vanishing expectation values automatically,
taking into account the        
above assumed list of the nonvanishing 2-point functions.  One should also note, that the auxiliary fields $X,Y^0 \sim {\cal{O}}(n)$.  
Comparing the solutions of the equations for the two auxiliary fields, one   finds a relation valid both in renormalized and unrenormalized     
versions:        
\be        
\sqrt{g_1}Y^0=\sqrt{g_2}X.        
\label{Eq:saddle-relation}
\ee        
Also, it proves useful in the further discussion to introduce the combination        
\be        
M^2=m^2-\frac{i}{n}(\sqrt{2g_1}X+\sqrt{2g_2}Y^0).        
\label{Eq:mass-gap}        
\ee        
          
\subsection{The propagator equations}        
        
The second derivatives of the effective action yield the various 2-point functions. 
It is convenient to introduce the following tree-level inverse propagator
commonly appearing in several two-point functions:        
\be        
iD_0^{-1}(p)=p^2-M^2.        
\ee
In this        
section we substitute for all 3-point functions their classical        
expressions listed below in ({\ref{Eq:bare-vertex}}) 
and therefore close the
coupled set of DS equations at the 2-point level (BVA).   
From ({\ref{Eq:shifted-aux-L}}) one can read off the nonzero
classical 3-point couplings:       
\bea        
&        
\displaystyle        
\Gamma_{X\pi^a\pi^b}=\Gamma_{Xs^as^b}=i\frac{\sqrt{2g_1}}{n}\delta^{ab}        
\delta(x-y)\delta(x-z),        
\nonumber\\        
&\displaystyle        
\Gamma_{Y^as^bs^c}=\Gamma_{Y^a\pi^b\pi^c}=i\sqrt{\frac{g_2}{n}}        
d^{abc}\delta(x-y)\delta(x-z),\qquad        
\Gamma_{Y^as^b\pi^c}=-i\sqrt{\frac{g_2}{n}}f^{abc}\delta(x-y)\delta(x-z).     
\label{Eq:bare-vertex}  
\eea        
      
In the pseudoscalar sector the following DS-equations are     
found at leading order (the abbreviated 
notation $\int G_{[.]}G_{[.]}\equiv \int_k G_{[.]}(k) G_{[.]}(p+k)$ is used):  
\bea        
iG_{\pi^0}^{-1}&=&iD_0^{-1}-2ig_2\int G_Y G_{\pi},\nonumber\\        
iG_{\pi}^{-1}&=&iD_0^{-1}-ig_2\int G_{Y}(G_s+G_{\pi})+ig_2\int G_{sY}G_{sY}.  
\label{Eq:pionprop}  
\eea     
One notes the potential violation of Goldstone's 
theorem when comparing       
$G_{\pi}(k=0)$ or $G_{\pi^0}(k=0)$ with the equation of state. Nevertheless,     
our approximate solution obeys this theorem due to a rather    
nontrivial relation  between the relevant tadpole and bubble contributions (see next section).      
The previous relations also hint to a 
possible further dynamical violation of the        
remaining $U(n)$ symmetry to $SU(n)$ due to the coupling of the auxiliary
 variables $Y^u$ to the        
scalar fields $s^u$, which is proportional to the antisymmetric      
structure tensor $f^{abc}$.           
This coupling can be seen explicitly in the structure of the $Y^a-s^a$ sector, where (if $a\neq 0$)        
one finds $n^2-1$ identical mixing $2\times 2$ DS equations:  
\bea        
&\displaystyle  
iG_{s}^{-1}=iD_0^{-1}-ig_2\int G_{sY}G_{sY}-ig_2\int G_{Y}(G_{s}+G_{\pi}), 
\qquad iG_{Y}^{-1}=-1-ig_2\int G_s G_{\pi}-i\frac{g_2}{2}\int(G_sG_s+G_{\pi}G_{\pi}), \nonumber\\        
&  
\displaystyle  
iG_{sY}^{-1}=2i\sqrt{g_2}v-ig_2\int G_{sY}(G_s-G_{\pi}).  
\label{Eq:sYprop}  
\eea         
In the $3\times 3$ mixing sector of $(X,Y^0,s^0)$, the following 6        
equations are obtained:        
\bea        
&\displaystyle  
iG_{X}^{-1}=-1-ig_1\int(G_sG_s+G_{\pi}G_{\pi}), \qquad iG_{XY^0}^{-1}=-i\sqrt{g_1g_2}\int(G_sG_s+G_{\pi}G_{\pi}), \nonumber\\        
&  
\displaystyle  
iG_{s^0X}^{-1}=2i\sqrt{g_1}v-2i\sqrt{g_1g_2}\int G_sG_{sY}, \qquad iG_{Y^0}^{-1}=-1-ig_2 \int(G_sG_s+G_{\pi}G_{\pi}), \nonumber\\      
&  
\displaystyle  
iG^{-1}_{s^0Y^0}=2i\sqrt{g_2}v-2ig_2\int G_s G_{sY}, \qquad iG^{-1}_{s^0}=iD_0^{-1}-2ig_2\int(G_{s}G_{Y}+G_{sY}G_{sY}).  
\label{Eq:s0XY0prop}  
\eea          
One can collect the previously derived propagator equations and formulate them as matrices:  
\be        
i{\cal G}_{(s,Y)}^{-1}=        
\begin{pmatrix}        
~iD_0^{-1}-ig_2\int[G_Y(G_s+G_{\pi})+G_{sY}G_{sY}]~  &        
~2i\sqrt{g_2}v-ig_2\int G_{sY}(G_s-G_{\pi})~\\        
~2i\sqrt{g_2}v-ig_2\int G_{sY}(G_s-G_{\pi})~        
&~-1-i\frac{g_2}{2}[{\cal A}+\int 2G_sG_{\pi}]~        
\end{pmatrix},      
\label{Eq:sy-sector}        
\ee        
\be        
i{\cal G}_{(X,Y^0,s^0)}^{-1}=        
\begin{pmatrix}        
-1-ig_1{\cal A}~  & ~-i\sqrt{g_1g_2}{\cal A}~ &        
~2i\sqrt{g_1}v-2i\sqrt{g_1g_2}\int G_{s}G_{sY}\\        
-i\sqrt{g_1g_2}{\cal A} & -1-ig_2{\cal A}~ &        
~2i\sqrt{g_2}v-2ig_2\int G_{s}G_{sY}\\        
2i\sqrt{g_1}v-2i\sqrt{g_1g_2}\int G_{s}G_{sY}~ &        
~2i\sqrt{g_2}v-2ig_2\int G_{s}G_{sY}~        
&~iD_0^{-1}-2ig_2\int (G_s G_Y+G_{sY}G_{sY})        
\end{pmatrix}.        
\label{Eq:s0y0x-sector}        
\ee          
Here we introduced the notation ${\cal A}=\int(G_sG_s+G_{\pi}G_{\pi})$.   
          
\subsection{Construction of the two-loop 2PI effective potential}

In this subsection a 2PI effective potential ($V^{2PI}$) is given, from which the previously determined
equations can be obtained directly by functional differentiation. The functional depends on the background $v$, the        
auxiliary (composite) fields ($X,Y^0$) and the 2-point     
functions. One should observe, that in (\ref{Eq:pionprop}), (\ref{Eq:sy-sector}) and (\ref{Eq:s0y0x-sector})
the corrections to the tree-level propagators are given by 1-loop diagrams. Such terms arise from 2-loop vacuum diagrams contributing to 
the 2PI effective potential by "cutting up" the lines corresponding to a propagator under functional differentiation
(for an introduction see \cite{berges}).
This means, that we have to draw 2-loop vacuum diagrams with the participation of the original and the auxiliary fields to reproduce
 the bubble contributions in the Dyson-Schwinger equations above (for the
pion propagator equation the process is demonstrated on Fig. \ref{Fig1}).
This construction therefore corresponds to a specific truncation of the 2PI effective
potential: from the infinite 2PI skeleton diagrams only a set of 2-loop graphs is taken into account.

\begin{figure}[!t]
\centerline{ 
\includegraphics[bb=369 612 356 699]{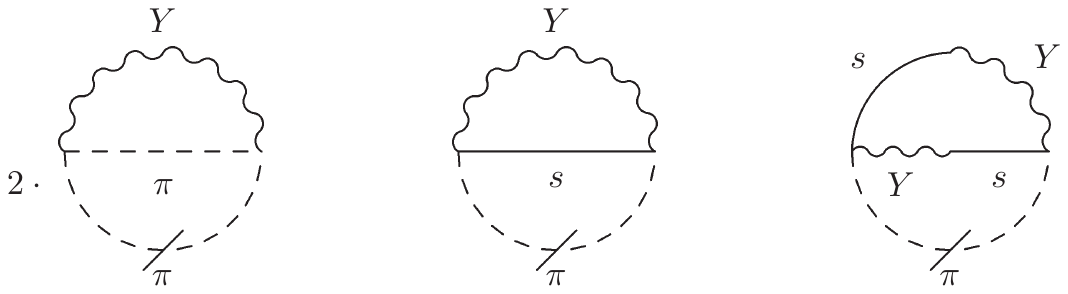}}
\caption{Vacuum diagrams contributing to the leading order equation of $G_\pi$ (cf. the ``setting sun'' integrals of (\ref{Eq:eff-pot-rescaled}) containing $G_\pi$ and proportional to $n^2$). One pion line is cut corresponding to the variation of the 2PI effective potential with respect to the pion propagator.}
\label{Fig1}
\end{figure}

The following zeroth order
(tree-level) propagator matrices from ({\ref{Eq:shifted-aux-L}}) will also appear in the 2PI effective potential:        
\be        
\displaystyle        
i{\cal D}_{(s,Y)}^{-1}=        
\begin{pmatrix}        
~iD_0^{-1}~ &        
~2i\sqrt{g_2}v~\\        
~2i\sqrt{g_2}v~        
 & ~-1~        
\end{pmatrix}, \qquad       
\displaystyle   
i{\cal D}_{(X,Y^0,s^0)}^{-1}=  
\begin{pmatrix}        
~-1~ & 0 & ~2i\sqrt{g_1}v~ \\        
0 & -1 & ~2i\sqrt{g_2}v~ \\        
 ~2i\sqrt{g_1}v~ & ~2i\sqrt{g_2}v~ & iD_0^{-1} \\       
\end{pmatrix}.        
\label{Eq:treelevel}        
\ee          
At this point it is convenient to introduce the rescaled variables    
\be    
x=\frac{\sqrt{2g_1}}{n}X,\qquad y^0=\frac{\sqrt{2g_2}}{n}Y^0,    
\ee    
which leads to $M^2=m^2-i(x+y^0)$. Using (\ref{Eq:1pointeq}),  
(\ref{Eq:pionprop}), (\ref{Eq:sy-sector}) and (\ref{Eq:s0y0x-sector}), one writes        
\bea        
V^{2PI}&=&n^2(m^2v^2-2hv)+\frac{n^2}{4}\left(\frac{1}{g_1}x^2+\frac{1}{g_2}    
(y^0)^2\right)-        
in^2(y^0+x)v^2\nonumber\\        
&&        
-i\frac{n^2}{2}\int_k(\ln G_{\pi}^{-1}(k)+D_0^{-1}(k)G_{\pi}(k))        
-i\frac{n^2}{2}\int_k\textrm{Tr}[\ln{\cal G}_{(s,Y)}^{-1}(k)+        
{\cal D}_{(s,Y)}^{-1}(k){\cal G}_{(s,Y)}(k)]             
\nonumber\\
&&        
-i\frac{1}{2}\int_k(\ln G_{\pi^0}^{-1}(k)+D_0^{-1}(k)G_{\pi^0}(k))        
-i\frac{1}{2}\int_k\textrm{Tr}[\ln{\cal G}_{(X,Y^0,s^0)}^{-1}(k)+        
{\cal D}_{(X,Y^0,s^0)}^{-1}(k){\cal G}_{(X,Y^0,s^0)}(k)]             
\nonumber\\
&&
+i\frac{n^2}{4}g_2\int_k\int_p\Big[\Big(G_{\pi}(k)G_{\pi}(k+p)+        
G_s(k)G_s(k+p)\Big)G_{Y}(p)+2G_{sY}(k)G_{sY}(k+p)G_s(p)\Big]        
\nonumber        
\\        
&&        
+i\frac{n^2}{2}g_2~\int_k\int_p\Big(G_{Y}(k)G_s(k+p)-G_{sY}(k)      
G_{sY}(k+p)\Big)G_{\pi}(p)
\nonumber
\\
&&
+i\frac{1}{2}g_1\int_k\int_p\Big(G_s(k)G_s(k+p)+G_{\pi}(k)G_{\pi}(k+p)\Big)G_{X}(p)
\nonumber
\\
&&
+i\sqrt{g_1g_2}\int_k\int_p\Big(G_s(k)G_s(k+p)+G_{\pi}(k)G_{\pi}(k+p)\Big)G_{XY^0}(p)
\nonumber
\\
&&
+i\frac{1}{2}g_2\int_k\int_p\Big(G_s(k)G_s(k+p)+G_{\pi}(k)G_{\pi}(k+p)\Big)G_{Y^0}(p)
\nonumber
\\
&&
+2i\sqrt{g_1g_2}\int_k\int_p G_{sY}(k)G_{s}(k+p)G_{s^0X}(p)+2ig_2\int_k\int_p G_{sY}(k)G_{s}(k+p)G_{s^0Y^0}(p)
\nonumber
\\
&&
+i\frac{1}{2}g_2\int_k\int_p\Big(G_s(k)G_Y(k+p)+G_{sY}(k)G_{sY}(k+p)\Big)G_{s^0}(p)
\nonumber
\\
&&+ig_2\int_k\int_p G_{Y}(k)G_{\pi}(k+p)G_{\pi^0}(p)+V^{2PI}_{ct}.        
\label{Eq:eff-pot-rescaled}        
\eea       
Since 
the $\pi$ and the $s-Y$ sectors are of multiplicity ($n^2-1$), some
${\cal{O}}(1)$ terms in (\ref{Eq:eff-pot-rescaled}) are not relevant for their leading order
equations. These contributions were therefore not written out
explicitly in (\ref{Eq:pionprop}) and (\ref{Eq:sYprop}). There are
${\cal{O}}(1)$ terms of the complete effective potential, which contribute only to the $\pi$ or $s-Y$ sector
(at NLO), these are omitted from (\ref{Eq:eff-pot-rescaled}).
However, for the remaining
propagators it is necessary to take into account 
the proper ${\cal{O}}(1)$ terms, which
 are therefore included in (\ref{Eq:eff-pot-rescaled}). This means, that
(\ref{Eq:eff-pot-rescaled}) is {\it{not}} a large-$n$ expanded 2PI
effective potential truncated at 2-loop level.
Note, that the structure appearing in the fifth line corresponds to a setting-sun      
diagram with antisymmetrized vertex functions (e.g. $\sim      
f^{abc}$). 
    
In $V^{2PI}_{ct}$ one collects        
all counterterms, which should ensure the finiteness of the equations. One can introduce counterterms to all
independent pieces appearing in the mean-field part of the effective potential and independently of them also to the 
terms showing up in the inverse tree-level propagators. Note, that one might have to introduce counterterms
also to pieces which are chosen to have a fixed numerical (zero or unity) renormalized coefficient. No counterterms are introduced corresponding to the setting-sun
contributions, since these couplings are kept at their classical value.
In this sense the most general form which we will need and is allowed by the structure of $V^{2PI}$ is the following:     
\bea    
V^{2PI}_{ct}&=&n^2\delta m^2_0v^2+\frac{n^2}{4}\Big(\delta_xx^2+\delta_y(y^0)^2\Big)    
+\frac{n^2}{2\sqrt{g_1g_2}}\delta_{xy}xy^0    
-in^2v^2(\delta_{yv}y^0+\delta_{xv}x)-in^2(\delta_{y0}y^0+\delta_{x0}x)    
\nonumber\\    
&&-i\int_k \left[\frac{n^2}{2}\delta_{YY}G_{Y}(k)+\delta_{XX}G_{X}(k)+    
\delta_{Y^0Y^0}G_{Y^0}(k)+2\delta_{XY^0}G_{XY^0}(k)\right]  
\nonumber\\  
&&-\frac{n^2}{2}\int_k\left[(\delta Z_\pi k^2-\delta
m^2_\pi+i\delta_{x\pi}x+i\delta_{y\pi}y^0)G_\pi(k)+
(\delta Z_s k^2-\delta
m^2_s+i\delta_{xs}x+i\delta_{ys}y^0)G_s(k)\right]   
\label{Eq:counterfunct}    
\eea    
(the self-energy corrections to $G_{sY}^{-1}$ turn out to be finite).
In the first line, countercouplings proportional to $xy^0, y^0, x$ will be also needed despite the fact that one chooses
the corresponding renormalized values to vanish. Similarly, countercouplings in the second line belong to the contribution
of the (pure) auxiliary propagators occurring in terms of the type ${\textrm {Tr}}{\cal D}^{-1}{\cal G}$, in which all their 
renormalized coefficients are fixed to unity except the coefficient of $G_{XY^0}$, which is zero. In the third line, countercouplings are introduced
to the terms appearing in $D^{-1}_{ss}$ and $D^{-1}_\pi$. In the exact solution of the theory only unique quadratic and quartic counterterms should occur,
but in any finite order of a 2PI approximation 
one has the freedom to choose the countercouplings in the counterterm functional
independently \cite{berges05}.
For the determination of (\ref{Eq:counterfunct}) one should analyze
the divergence structure of the integrals appearing in the propagator equations.  
        
\section{A self-consistent assumption on mass hierarchy}        
          
The assumption, that the scalar sector is considerably heavier than      
the pionic simplifies the solution of the        
limiting form of the coupled Dyson-Schwinger equations valid at        
infinite $n$. In addition, we assume that the scalar masses are      
considerably larger than the amplitude of the symmetry breaking vacuum      
condensate.  Each component of the propagators in the mixed sectors will have a common
denominator displaying the corresponding heavy mass. Therefore the only consequent way to neglect in a first approximation
the heavy sector is to neglect the bubble contributions containing at least one component of ${\cal G}_{(s,Y)}$ or ${\cal G}_{(X,Y^0,s^0)}$.
Then, in a first approximation only bubble diagrams exclusively
built with $G_{\pi}$  are included. 
 This means that in (\ref{Eq:pionprop}) all bubbles are suppressed, while in (\ref{Eq:sy-sector}) and (\ref{Eq:s0y0x-sector}) only the unique pure pion bubble remains.
      
\subsection{Retaining only the light pion dynamics} 
       
The first consequence is that all $n^2$ pion propagators are equal to their tree-level value, and therefore have the same mass: $M^2_{\pi^a}=M^2$.        
The explicit form of the saddle-point equations can be written        
 with the help of splitting the pion tadpole into finite $(T_\pi^F)$ and 
divergent parts [for the definitions of $T_d^{(2)}, T_d^{(0)}$ see Appendix A.] as     
\be     
\int_k G_{\pi}(k)=T_\pi^F+T_d^{(2)}+(M^2-M_0^2)T_d^{(0)},      
\ee  
and using the counterterms it becomes the following:    
\bea        
\frac{x}{g_1}=i\Big(2v^2+T_d^{(2)}+(M^2-M_0^2)T_d^{(0)}+T_\pi^F\Big)-   
\Big(\delta_xx+\delta_{xy}\frac{1}{\sqrt{g_1g_2}}y^0\Big)+2i\delta_{x0},        
\nonumber\\         
\frac{y^0}{g_2}=    
i\Big(2v^2+T_d^{(2)}+(M^2-M_0^2)T_d^{(0)}+T_\pi^F\Big)    
-\Big(\delta_yy^0    
+\delta_{xy}\frac{1}{\sqrt{g_1g_2}}x\Big)+2i\delta_{y0}.        
\label{Eq:xeq_ren}  
\eea       
It turns out, that at this level the other counterterms (e.g. $\delta_{xv},\delta_{x\pi},\delta_{xs},\delta_{yv},
\delta_{y\pi},\delta_{ys}$) are zero, therefore for the
sake of transparency we did not written them out explicitly in (\ref{Eq:xeq_ren}).    
Taking into account the definition (\ref{Eq:mass-gap}) of $M^2$,        
the following counterterms renormalize both equations:        
\be        
\delta_x=\delta_y=T_d^{(0)},\qquad    
\delta_{x0}=\delta_{y0}=-\frac12\left(T_d^{(2)}+(m^2-M_0^2)T_d^{(0)}\right),\qquad    
\delta_{xy}=\sqrt{g_1g_2}T_d^{(0)}.       
\label{Eq:cterms-one-point}        
\ee        
Because of the omission of the last ("heavy") term in the third equation of ({\ref{Eq:1pointeq}}), 
the equation of state does not require any extra counterterm.        
The finite equations for the 1-point functions read as follows:        
\be        
\frac{x}{g_1}=i(2v^2+T_\pi^F),        
\qquad \frac{y^0}{g_2}=i(2v^2+T_\pi^F),\qquad M^2v=h.    
\label{Eq:1-point-eqs}  
\ee          
The $2\times 2$ propagator matrix of the $(s,Y)$ sector simplifies to        
\be        
\displaystyle        
i{\cal G}_{(s,Y)}^{-1}(k)=        
\begin{pmatrix}        
~iD_0^{-1}(k)~ &        
~2i\sqrt{g_2}v~\\       
~2i\sqrt{g_2}v~        
 & ~-1+i\delta_{YY}+\frac{g_2}{2}I_{\pi}(k,M)~        
\end{pmatrix},        
\label{Eq:sy-sector-noscalar}        
\ee        
where we have introduced the function $I_{\pi}(k,M)=-i\int_pD_0(p)D_0(p+k)$ and also    
written down the contribution obtained from $V^{2PI}_{ct}$.    
By choosing    
\be    
\delta_{YY}=i\frac{g_2}{2}T_d^{(0)},
\ee    
one ensures the finiteness of the matrix elements.        
The squared scalar mass is determined by the zero of the determinant:        
\be        
\displaystyle        
M_{(s,Y)}^2=M^2+4g_2v^2\frac{1}{1-\frac{g_2}{2}I_{\pi}^F(k=M_{(s,Y)},M)}.    
\label{Eq:sYgap}    
\ee      
        
The mass matrix of the $(X,Y^0,s^0)$ sector also becomes more transparent:        
\be        
i{\cal G}_{(X,Y^0,s^0)}^{-1}=        
\begin{pmatrix}        
~-1+2i\delta_{XX}+g_1I_{\pi}(k)~  & ~2i\delta_{XY^0}    
+\sqrt{g_1g_2}I_{\pi}(k)~ &        
~2i\sqrt{g_1}v~\\        
 ~2i\delta_{XY^0}+\sqrt{g_1g_2}I_{\pi}(k)~& ~-1+2i\delta_{Y^0Y^0}    
+g_2I_{\pi}(k)~ &        
~2i\sqrt{g_2}v~\\        
~2i\sqrt{g_1}v~& ~2i\sqrt{g_2}v~        
& ~iD_0^{-1}(k)~        
\end{pmatrix},        
\label{Eq:s0y0x-sector-noscalar}        
\ee        
which after introducing the obvious counterterms    
\be    
\delta_{XX}=\frac{i}{2}g_1T_d^{(0)},\qquad \delta_{XY^0}=\frac{i}{2}\sqrt{g_1g_2}T_d^{(0)},    
\qquad \delta_{Y^0Y^0}=\frac{i}{2}g_2T_d^{(0)}    
\ee    
leads to a determinant equation completely analogous to the        
previous one:        
\be        
\displaystyle        
M_{(X,Y^0,s^0)}^2=M^2+4(g_1+g_2)v^2\frac{1}{1-(g_1+g_2)I^F_{\pi}(k=M_{(X,Y^0,s^0)},M)}.  
\label{Eq:xY0s0gap}       
\ee        
One can see, that the spectra resulting from the assumption we made for        
the mass hierarchy might be consistent with the outcome in the sense that      
the scalar sector (and the auxiliary fields hybridized with it) can be  
indeed heavier than the pionic one.       
      
In the next subsection we investigate the region of the parameter space, where the
scalar masses become much heavier than the pseudoscalars
ensuring the self-consistency of our approximate solution.        
      
\subsection{Validity of the mass assumption}

First one has to note, that in the case when no explicit symmetry breaking term is
added to the Lagrangian ($h=0$) the mass assumption is true, since our approximation
preserves Goldstone's theorem and therefore makes the pions massless, which means that they
are "infinitely" lighter than the scalars.

The interesting case is when $h\neq 0$. Let us define $r_{s,Y}:=M_{(s,Y)}/M$ and $r_{X,Y^0,s^0}:=M_{(X,Y^0,s^0)}/M$. In order to obtain a proper region of the 
parameter space, we introduce a heaviness criterium: the scalars are heavy enough if relations $r_{s,Y}>r_0$, $r_{X,Y^0,s^0}>r_0$ hold simultaneously, where $r_0$ is a given
number.  
It is somewhat arbitrary what value to choose for $r_0$. In this exploratory study, we work with the convenient choice $r_0=2$, because the quantity $I_\pi^F(k,M)$ appearing in both gap equations ({\ref{Eq:sYgap}) and ({\ref{Eq:xY0s0gap}) develops an imaginary part just for $k^2>4M^2$ (two-pion threshold). Above the threshold the masses are defined as real parts of the complex solutions. Expressing $v$ from the EoS in ({\ref{Eq:1-point-eqs}), the two relevant equations are the following:

\bea
M_{(s,Y)}^2&=&M^2+4g_2\frac{h^2}{M^4}\frac{1-\frac{g_2}{2}\Re I_{\pi}^F(k=M_{(s,Y)},M)}{[1-\frac{g_2}{2}\Re I_{\pi}^F(k=M_{(s,Y)},M)]^2+[\frac{g_2}{2}\Im I_{\pi}^F(k=M_{(s,Y)},M)]^2}, \\
M_{(X,Y^0,s^0)}^2&=&M^2+4(g_1+g_2)\frac{h^2}{M^4}\frac{1-(g_1+g_2)\Re I_{\pi}^F(k=M_{(X,Y^0,s^0)},M)}{[1-(g_1+g_2)\Re I_{\pi}^F(k=M_{(X,Y^0,s^0)},M)]^2+[(g_1+g_2)\Im I_{\pi}^F(k=M_{(X,Y^0,s^0)},M)]^2}. \nonumber
\label{Eq:scalar-gaps}
\eea  
The bubble integrals are given by ({\ref{Eq:fin-bub}}). It is convenient to express all masses in proportion to the absolute value of the renormalized mass $(\sqrt{|m^2|})$, which means practically to write the definition of the pion mass as $M^2=-1-ix-iy_0$. With the use of ({\ref{Eq:1-point-eqs}) and ({\ref{Eq:TF}}), one gets
\be
M^2=-1+(g_1+g_2)\Big(\frac{2h^2}{M^4}+\frac{M^2}{16\pi^2}\log \frac{eM^2}{M_0^2}\Big),
\label{Eq:pion-gap}
\ee
where $M_0$ is the renormalization scale. For fixed $M_0$, in the original units ({\ref{Eq:pion-gap}}) determines
$M^2/|m^2|$ as a function of $h/|m|^3,g_1,g_2$. 
Plugging it into the scalar gap equations, they
can be solved for $M_{(s,Y)}/\sqrt{|m^2|}$ and $M_{(X,Y^0,s^0)}/\sqrt{|m^2|}$. Then, one can trace out the region, where the heaviness criterium is fulfilled. This region is the part of the positive $g_1-g_2$ octant (the stability region of the $n\rightarrow\infty$ theory) below the surface displayed in Fig. 2. As expected, the projection of the allowed region onto the $g_1-g_2$ plane shrinks for increasing value of $h/|m|^3$. We have varied $M_0/\sqrt{|m|^2}$ in the interval $(1,10)$ and only a mild displacement of the allowed region could be observed without really changing the shape. This change can be balanced by an appropriate Renormalization Group transformation of the quartic couplings, e.g. using $g_i=g_i(M_0)$.

The modification of the allowed region occuring when the spectra is corrected by the fluctuations of the heavy scalars will be discussed in a forthcoming publication.

\begin{figure}[!t]
\centerline{ 
\includegraphics[bb=124 72 300 279,scale=0.81]{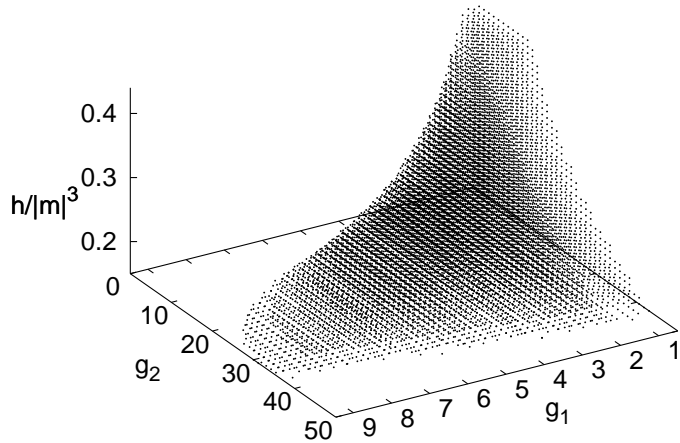}}
\caption{Region of the parameter space, where the mass assumption holds (using renormalization scale $M_0/\sqrt{|m^2|}=2.5$)}
\label{Fig2}
\end{figure}

\subsection{Heavy scalar corrections} 
     
 Using the propagators of the coupled $(s,Y)$-sector, one can     
 start to systematically take into account the effect of the heavy      
 degrees of freedom on the pion propagator, the EoS and the SPEs. 
 In particular we now invoke the tadpole and bubble contributions to these equations
evaluated with
 ({\ref{Eq:sy-sector-noscalar}}).   
For this we write explicitly the components of the heavy scalar      
propagator matrix:    
\bea      
&\displaystyle      
G_{s}(k)=\frac{i}{d(k)}\left(1-\frac{g_2}{2}I_{\pi}^F(k)\right),      
\quad G_{Y}(k)=-\frac{i}{d(k)}(k^2-M^2),\quad      
G_{sY}(k)=-\frac{1}{d(k)}2\sqrt{g_2}v,\nonumber\\      
&\displaystyle      
d(k)=\left(1-\frac{g_2}{2}I_{\pi}^F(k)\right)(k^2-M^2)-4g_2v^2.      
\eea

With the help of these expressions, one readily writes the scalar tadpole      
contributions to the SPEs.       
The correction of the EoS is of imminent interest, since it has      
an important role in the discussion of the validity of Goldstone's      
theorem. Using the expression of $G_{sY}$, one finds the unrenormalized equation of state:      
\be\displaystyle      
M^2+2g_2\int_p\frac{i}{d(p)}=\frac{h}{v}.   
\ee      
Now we proceed with the unrenormalized pion propagators:      
\bea      
iG_{\pi^0}^{-1}(k)&=&k^2-M^2-2g_2\int_p\frac{i}{d(p)}\frac{p^2-M^2}    
{(k-p)^2-M^2},      
\nonumber\\      
iG_{\pi}^{-1}(k)&=&k^2-M^2-g_2\int_p\frac{i}{d(p)}      
\frac{p^2-M^2}{(k-p)^2-M^2}\nonumber\\      
&&+4g_2^2v^2\int_p\frac{i}{d(p)d(k-p)}-ig_2\int_p      
\frac{(p^2-M^2)(1-g_2I_{\pi}^F(k-p)/2)}{d(p)d(k-p)}.      
\label{Eq:pion-props}
\eea      
When one sets $k=0$, both propagator equations turn out to be the  
same. Comparing to the equation of state one  
finds, that the approximation fulfills the Goldstone theorem
characterizing the $U(n)\times U(n)\rightarrow U(n)$ symmetry
breaking. It will be demonstrated in the next section, that both
equations receive the same counterterm contributions.
 
Since the tadpoles of $G_s$ and $G_{sY}$ play a role in the EoS and
the SPEs, it is worthwhile to investigate their corrections, although
these contributions should be considered only as NNLO ``heavy'' corrections.
Substituting the leading order propagators into the second term on the
right hand side of the equation of $G_{sY}^{-1}$ (\ref{Eq:sYprop}), one
finds for the $\Sigma_{sY}$ self-energy:
\be
\Sigma_{sY}(p)\equiv ig_2\int_k G_{sY}(p-k)\Big(G_s(k)-G_{\pi}(k)\Big)=-8g_2^2\sqrt{g_2}v^3\int_k\frac{1}{d(p-k)d(k)(k^2-M^2)},
\ee 
which does not induce any counterterm, since it is finite.          
The heavy correction to $G_s^{-1}$ is very similar to that of $G^{-1}_{\pi}$ in (\ref{Eq:pion-props}):
\bea         
iG_{s}^{-1}(k)&=&k^2-M^2-g_2\int_p\frac{i}{d(p)}      
\frac{p^2-M^2}{(k-p)^2-M^2}\nonumber\\      
&&-4g_2^2v^2\int_p\frac{i}{d(p)d(k-p)}-ig_2\int_p      
\frac{(p^2-M^2)(1-g_2I_{\pi}^F(k-p)/2)}{d(p)d(k-p)}.      
\label{Eq:s-prop}
\eea 

The scalar corrections are taken into account also in the SPEs. They appear partly directly via the scalar tadpole, 
and also by the scalar correction of the pion tadpole. The structure of    
the two SPE's in ({\ref{Eq:1pointeq}}) is identical in view of
(\ref{Eq:saddle-relation}), therefore it is sufficient to investigate
the equation, which determines $x$. The unrenormalized form of the
saddle-point equation reads as follows:    
\begin{equation}    
0=\frac{x}{g_1}-2iv^2-i\int_k\left(G_s(k)+G_{\pi}(k)\right).  
\label{Eq:ren-spex}    
\end{equation}
The expression of the scalar tadpole is readily written down.           
The pion tadpole is expanded to linear order in the $\Sigma_{\pi}$ self-energy contribution:
\be
-i\int_kG_\pi(k)\approx -i\int_kD_0(k)+\int_k\frac{1}{(k^2-M^2)^2}\Sigma_{\pi},
\label{Eq:pion_prop-corr}
\ee
where
\be
\Sigma_\pi(k)=g_2\int_p\frac{i}{d(p)}\left[\frac{p^2-M^2}
{(k-p)^2-M^2}
-\frac{4g_2v^2}{d(k-p)}+\frac{(p^2-M^2)(1-g_2I^F_\pi(k-p)/2)}{d(k-p)}\right].
\ee

All these equations need (resummed) renormalization, which is discussed in detail in the next section.
Note, that in this process the counterterms (\ref{Eq:cterms-one-point}), which were determined
on the level of pure pionic fluctuations receive further contributions. Also, the yet unused counterterms (except
$\delta Z_s, \delta Z_\pi$ counterterms proportional to wave function renormalization) will get nonzero values.

\section{Divergence analysis and renormalizability}      
      
The separation of the divergences in the EoS, the corrections to the
pion propagator and the SPEs rely very heavily
 on our previous analysis of the NLO  
renormalization of the $O(N)$ model \cite{fejos09}. Most of the
divergent integrals occurring in the present analysis 
can be made finite with the subtraction of the appropriate
combinations of divergent integrals defined there. For the reader's
convenience we list those, which are used here in Appendix A.    
      
Let us start the divergence analysis with the EoS. The integral appearing 
in it can be expanded for large momenta in powers of $4g_2v^2$:      
\be      
\int_p\frac{i}{d(p)}=\int_p\frac{i}{(p^2-M^2)(1-g_2I_\pi^F(p)/2)}\left(      
1+\frac{4g_2v^2}{(p^2-M^2)(1-g_2I_\pi^F(p)/2)}+ ...\right).      
\ee      
Only the first term of this expansion is divergent and its divergence      
is given in (\ref{Eq:div-i1}) with $\gamma=g_2/2$:      
\be      
\int_p\frac{i}{d(p)}\Big|_{div}=T_a^{(2)}-\frac{3g_2}{2}(M^2-M_0^2)T_a^{(I)}.  
\ee      
From here one finds for the corresponding countercouplings of
(\ref{Eq:counterfunct}):      
\be      
\delta      
m_0^2=-2g_2\left(T_a^{(2)}-\frac{3g_2}{2}(m^2-M_0^2)T_a^{(I)}\right),      
\qquad \delta_{xv}=\delta_{yv}=3g_2^2T_a^{(I)}.      
\ee      
      
Next we proceed with the $\Sigma_{\pi}$ self-energy. 
The second of the three      
bubble integrals appearing in the expression of the heavy scalar      
corrections to $G^{-1}_\pi$ (\ref{Eq:pion-props}) is finite. 
The first integral can be      
written in power series with respect to $4g_2v^2$ as      
\bea      
&&-ig_2\int_p\frac{1}{d(p)}\frac{p^2-M^2}{(k-p)^2-M^2}\nonumber\\      
&&=      
-ig_2\int_p\frac{1}{(p^2-M^2)(1-g_2I_\pi^F(p)/2)}\left(1+\frac{4g_2v^2}      
{(p^2-M^2)(1-g_2I_\pi^F(p)/2)}+ ...\right)\frac{p^2-M^2}{(k-p)^2-M^2}.      
\eea      
One recognizes again, that only the integral of the first term of the      
expansion is divergent. This is exactly the integral which was shown      
in Eq. (A7) of \cite{fejos09} to not have momentum dependent
divergence. The same analysis leads also in case of the third integral
to the same divergent piece.     
Therefore one concludes $\delta Z_\pi=0$.      
      
In view of this, one can put $k=0$ in the pion propagator and find its      
divergences. Since Goldstone's theorem is obeyed, the same      
divergences occur as in the EoS. As a consequence one has in
(\ref{Eq:counterfunct})      
\be      
\delta m_\pi^2=\delta m_0^2,\qquad \delta_{x\pi}=\delta_{xv},\quad      
\delta_{y\pi}=\delta_{yv}.      
\ee      
Using these results, one can find promptly the counterterms proportional
to the $G_s$ tadpole in (\ref{Eq:counterfunct}). Since the divergent
pieces of $\Sigma_s$ coincide with those appearing in $\Sigma_\pi$, one has
\be
\delta Z_s=0,\qquad \delta m_s^2=\delta m_0^2,\qquad
\delta_{xs}=\delta_{xv},
\qquad \delta_{ys}=\delta_{yv}.
\ee    
These equalities reflect the fact, that the ultraviolet behavior of the
$\pi$ and $s$ multiplets are the same, irrespective of the symmetry
breaking pattern and the structure of the mass spectrum.

At this stage we have specified the expressions of the    
countercouplings of the counterterm functional (\ref{Eq:counterfunct}), up
to the scalar corrections of the purely $X,Y^0$-dependent divergences.     
The renormalized SPE of $x$ takes a very transparent form, when the
results of the above analysis are taken into account:
\be
0=\frac{x}{g_1}-2i\left(v^2+\frac{1}{2}\int_k(G_s(k)+G_\pi(k))\right)
(1+\delta_{xv})+\delta_xx-2i\delta_{x0}+\frac{1}{\sqrt{g_1g_2}}\delta_{xy}y^0.
\label{Eq:ren-SPE}
\ee
The last three terms represent the counterterms, which cancel
those divergences of the tadpole integrals, which depend
solely on $x$ and $y^0$. The consistency with the previous steps of the renormalization
 requires the cancellation of the ``dangerous'' product
$\delta_{xv}(2v^2+\int G_s|_{F}+\int G_\pi|_F)\equiv\delta_{xv}(2v^2+T_s^F+T_\pi^F)$ without the need for any new counterterm (which would react back on the EoS and the equation of the pion's 2-point function).
 The unique common divergent
coefficient in front of this sum is expected to prevail in the exact renormalized equation, 
since it is the result of the Goldstone theorem and of the
 $U(n)\times U(n)$ symmetry valid in the ultraviolet regime.
 
The contributions cancelling the above counterterm contribution come
from the heavy corrections to the pion tadpole and from the scalar tadpole eventually also modified by the heavy corrections.    
This cancellation can not be complete on the level
of the first round of heavy corrections, 
since then the scalar tadpole does
not give any divergent term proportional to $T_s^F$ which would be
able to cancel $\delta_{xv}T_s^F$. However, this latter should be invoked first in the second iteration of the heavy corrections, therefore
one simply omits it at this stage of the iteration, and no imbalance is detected in the divergences proportional to $T_s^F$.  

The actual form of the SPE (\ref{Eq:ren-SPE}) is the following after  the first round of the heavy corrections:
\be
0=\frac{x}{g_1}-2i\left(v^2+\frac{1}{2}\int_k(G_s(k)+G_{\pi}(k))\right)
-2i\left(v^2+\frac{1}{2}\int_k D_0(k)\right)\delta_{xv}
+\delta_xx-2i\delta_{x0}+\frac{1}{\sqrt{g_1g_2}}\delta_{xy}y^0.
\label{Eq:ren-SPE-simpl}
\ee
Here one has to consider the pion tadpole as the sum displayed in
 (\ref{Eq:pion_prop-corr}).
The contribution of the scalar tadpole 
when expanded in powers of $4g_2v^2$ can be written as follows:    
\begin{equation}    
-i\int_kG_s(k)=-i\int_k\frac{i}{k^2-M^2}\left(1+4g_2v^2\frac{1}{(k^2-M^2)    
(1-g_2I^F_\pi(k)/2)}+...\right)    
\end{equation}    
Only these two terms, written out explicitly, contain divergences, 
therefore    
\begin{eqnarray}    
-i\int_kG_s(k)\Big|_{div}&=&-i[T_d^{(2)}+(M^2-M_0^2)T_d^{(0)}]
-4ig_2v^2\int_k\frac{i}{(k^2-M^2)^2  
(1-g_2I^F_\pi(k)/2)}\Big|_{div}\nonumber\\  
&=&-i[T_d^{(2)}+(M^2-M_0^2)T_d^{(0)}]+2ig_2^2v^2T_a^{(I)}.    
\label{scalar-tadpole-div}
\end{eqnarray}    
Every term, except the last one on the right-hand side can be cancelled 
by appropriate extensions of the counterterms 
at most quadratic in $x$ and $y^0$ [see (\ref{Eq:cterms-one-point})].
The heavy correction of the pion tadpole has the following expression:    
\begin{eqnarray}    
\Delta\left[-i\int_kG_{\pi}(k)\right]&=&-ig_2\int_k\frac{i}{(k^2-M^2)^2}
\times    
\nonumber\\&&\left(\int_p    
\frac{i}{d(p)}\left(\frac{p^2-M^2}{(k-p)^2-M^2}-\frac{4g_2v^2}{d(k-p)}      
+\frac{(p^2-M^2)(1-g_2I^F_\pi(k-p)/2)}{d(k-p)}\right)\right)_{finite}.    
\label{Eq:pion-tadpole-correction}  
\end{eqnarray}  
The part of the divergence which is independent of $v^2$ is separated as  
\begin{equation}  
\Delta \left[-i\int_kG_{\pi}(k)\right]\Bigg|_{v=0,div}=2ig_2  
\left[\int_k\frac{1}{(k^2-M^2)^2}\left(\int_p\frac{1}{((k-p)^2-M^2)(1-g_2I_{\pi}^F(p,M)/2)}\right)_{finite}\right]_{div}  
\end{equation}  
Using (\ref{Eq:div-i3}) of Appendix A., one finds  
\begin{equation}  
\Delta\left[-i\int_kG_{\pi}(k)\right]
\Bigg|_{v=0,~div}=-2ig_2(A_{div}+B_{div}M^2)+3ig_2^2  
T_\pi^F(M)T_a^{(I)}.  
\end{equation}  
The last term when substituted into (\ref{Eq:ren-SPE-simpl}) exactly compensates  
the dangerous term $-i\delta_{xv}T^F_\pi(M)$, which arises from the $-i\delta_{xv}\int D_0$
term. The appearing new divergences (proportional to $A_{div}$
and $B_{div}$) can be compensated
by adding further heavy corrections to (\ref{Eq:cterms-one-point}). The consistent cancellation of the divergences
proportional to $T_\pi^F$ could have been anticipated, since the counterterm
proportional to the scalar
tadpole left out from (\ref{Eq:ren-SPE}) at this level 
cannot lead to any divergence proportional to $T_\pi^F$. Therefore in this respect (\ref{Eq:ren-SPE-simpl}) does not differ from the exact relation
(\ref{Eq:ren-SPE}).

Next, one turns to the prospectively divergent terms proportional to $v^2$ of the SPE contributed by the pion tadpole:  
\begin{eqnarray} 
-4ig_2^2v^2\int_k&&\frac{i}{(k^2-M^2)^2}\times\nonumber\\  
&&\Bigl[2\int_p\frac{i}{(p^2-M^2)((k-p)^2-M^2)(1-g_2I^F_\pi(p)/2)^2}\nonumber\\  
&&  
+\int_p\frac{i}{((k-p)^2-M^2)(1-g_2I_\pi^F(k-p)/2)(1-g_2I^F_\pi(p)/2)}  
\left(\frac{1}{(k-p)^2-M^2}-\frac{1}{p^2-M^2}\right)\Bigr].  
\label{Eq:prop-v2}
\end{eqnarray} 
One finds by inspection, that
there is no subdivergence in this expression. 
The first term of the square bracket produces an overall divergent piece $8ig_2^2v^2T_a^{(I)}$. It can 
be found following the previous line of analysis, e.g. after changing the order of integration performing the 
$k$-integration first. A careful, but rather lengthy 
analysis leads to the conclusion, that
the second double integral is finite. 

Eventually one finds that in the sum the three divergent contributions proportional to $v^2$ do not annihilate. 
It is easy to see, at least partially, the source of this imbalance. 
When in a subsequent iteration round one uses in the scalar tadpole a propagator involving heavy corrections, 
it also produces a divergent contribution proportional to $ig_2^2v^2T_a^{(I)}$. This shows that in case of
 this single type of divergence for the cancellation one has to take into account contributions belonging to 
different iteration levels of the heavy contributions. Actually, all this is not
unexpected, since the ultraviolet features of the integrals are independent
of any hierarchy in the spectra.      
Therefore in practice, one renormalizes the saddle-point equations subtracting  the remainder 
of the divergences proportional to $v^2$ by hand.

\section{Conclusions and outlook}

A leading large-$n$ solution was presented in a ground state, which breaks 
the $U(n)\times U(n)$ symmetry of the Lagrangian to $U(n)$. In the
construction of the solution, a light pseudoscalar/heavy scalar
hierarchy of the spectra
was assumed. The renormalizability of the
equation of state and propagator equations, is a precondition for the investigation of the
consistency of this additional assumption. This feature was
demonstrated by constructing the counterterms to the equations above. The consistency 
of these counterterms required an additional subtraction for the renormalization of the saddle-point equations. It
was shown, that the proposed solution explicitly fulfills
Goldstone's theorem. The pions are generically light, therefore the assumed
mass hierarchy will not be erased by heavy radiative corrections,
as was demonstrated by our exploratory investigation.
It is probably present in a large part of the $(g_1,g_2)$
coupling-plane.

The proposed procedure of constructing a leading large-$n$ solution for the
$U(n)\times U(n)\rightarrow U(n)$ symmetry breaking pattern can be developed further in several directions.
In the range of validity of the mass hierarchy we shall study the finite temperature features of the proposed solution. The relevance of it would be largely strengthened, if a first order symmetry
restoring transition were found in agreement with the
renormalization group argument. For the applications to strong
interaction phenomena at $n=3$ one ought to introduce also the $U_A(1)$
breaking effective (determinant) 
term as a sort of perturbation to this solution. A simple realization could be to include its contribution 
into the two-loop 2PI effective potential of the model and evaluate it 
with approximate large-$n$ propagators constructed in this paper. Finally, we note, that 
one can make use of these propagators also in models, where constituent quarks are coupled to the mesons.

\begin{acknowledgments}
The authors are grateful to A. Jakov\'ac and Zs. Sz{\'e}p for many valuable suggestions.
This work is supported by the Hungarian Research Fund under Contracts
No. T068108 and K77534. 
\end{acknowledgments}

\makeatletter
\@addtoreset{equation}{section}
\makeatother

\renewcommand{\theequation}{A\arabic{equation}}

\appendix 
\section{Divergences of some relevant integrals}  
The divergences of the integrals listed below all can be read in somewhat scattered way in  \cite{fejos09}. 
Here we summarize them for the reader's convenience. The divergences  
are expressed in terms of the following divergent integrals:  
\begin{eqnarray}  
&\displaystyle  
T_d^{(2)}\equiv\int_k\frac{i}{k^2-M_0^2},\qquad   
T_d^{(0)}\equiv\int_k\frac{i}{(k^2-M^2_0)^2}\nonumber\\  
&\displaystyle  
T_a^{(2)}\equiv\int_k\frac{i}{(k^2-M_0^2)(1-\gamma I_0^F(k))},  
\qquad  
T_a^{(I)}\equiv\int_k\frac{i}{(k^2-M_0^2)^2(1-\gamma I_0^F(k))^2}I_0^F(k),  
\end{eqnarray}  
where $M_0$ is an arbitrary normalization scale which makes these integrals infrared safe, 
$\gamma$ is a parameter which in case of our model equals $g_2/2$, and $I^F(k,M)$ is
the finite part of the bubble integral  
\begin{equation}  
I^F(k,M)=\int_p\frac{i}{(p^2-M^2)((k-p)^2-M^2)}-T_d^{(0)}.  
\end{equation}  
or in more explicit terms with real and imaginary parts separated:
\bea
\Re I^F(k,M)&=& \frac{1}{16\pi^2}\log \frac{M^2}{M_0^2}-\frac{1}{16\pi^2} 
\begin{cases}
\sqrt{1-\frac{4M^2}{k^2}}\log \frac{1-\sqrt{1-\frac{4M^2}{k^2}}}{1+\sqrt{1-\frac{4M^2}{k^2}}}, \qquad \qquad \qquad k^2 \geq 4M^2 \\
-2\sqrt{\frac{4M^2}{k^2}-1}\arctan [1/\sqrt{1-\frac{4M^2}{k^2}}], \qquad 4M^2>k^2>0 \\
\sqrt{1-\frac{4M^2}{k^2}}\log \frac{\sqrt{1-\frac{4M^2}{k^2}}-1}{\sqrt{1-\frac{4M^2}{k^2}}+1}, \qquad \qquad \qquad 0 \geq k^2 \\
\end{cases}
\nonumber\\
\Im I^F(k,M)&=& -\frac{1}{16\pi}\sqrt{1-\frac{4M^2}{k^2}}\cdot \Theta(k^2-4M^2).
\label{Eq:fin-bub}
\eea
The finite part of the tadpole integral is
\begin{equation}
T^F(M)=\int_k \frac{i}{k^2-M^2}-(M^2-M_0^2)T_d^{(0)}-T_d^{(2)} \equiv
\frac{M^2}{16 \pi^2}\log \frac{M^2}{M_0^2}-\frac{M^2-M_0^2}{16 \pi^2}
\label{Eq:TF}
\end{equation}
  
The divergences of the integrals below were found by replacing
the mass parameter $M$ sequentially (with
help of subtractions and additions) by the normalization scale $M_0$:  
\begin{equation}  
I_1=\int_k\frac{i}{(k^2-M^2)(1-\gamma I^F(k,M))}\Big|_{div}=T_a^{(2)}-3\gamma  
(M^2-M_0^2)T_a^{(I)}\equiv \tilde T_{div}(M^2),  
\label{Eq:div-i1}  
\end{equation}  
where $I^F(k,M)$ is the bubble integral defined with mass $M$.  
We need in our analysis also the slightly modified integral:  
\begin{equation}  
I_{1a}=\int_k\frac{i}{(k^2-4M^2)(1-\gamma I^F(k,M))}\Big|_{div}=\tilde   
T_{div}(M^2)-3\gamma M^2T_a^{(I)}.  
\end{equation}  
There is a logarithmically divergent integral:  
\begin{equation}  
I_2=\int_k\frac{i}{(k^2-M^2)^2(1-\gamma I^F(k,M))}\Big|_{div}=-\gamma T_a^{(I)}.  
\label{Eq:div_i2}  
\end{equation}  
  
The most challenging is the separation of the setting-sun integral,  
where one subdivergence is already explicitly found:  
\begin{equation}  
I_3=\left[-\int_k\frac{i}{(k^2-M^2)^2}\left(\int_p\frac{1}{((k-p)^2-M^2)(1-\gamma I^F(p,M))}\right)_{finite}\right]_{div}.  
\end{equation}  
Using $I_1$ introduced above one can write  
\begin{equation}  
I_3=\left[-\int_k\frac{i}{(k^2-M^2)^2}\left(\int_p\frac{1}{((k-p)^2-M^2)(1-\gamma I^F(p,M))}+i\tilde T_{div}(M^2)\right)\right]_{div}.  
\end{equation}  
After changing the order of the integrals, using (A19) of \cite{fejos09}, one is led gradually to the final form of its divergences  
\begin{eqnarray}  
I_3&=&-\frac{1}{\gamma}\int_k\frac{1}{k^2-4M^2}\Big|_{div}+\left(\frac{1}{\gamma}-I^F(0,M)+  
\frac{1}{8\pi^2}\right)\int_k\frac{1}{(k^2-4M^2)(1-\gamma I^F(k,M))}  
\Big|_{div}\nonumber\\  
&&  
-i\tilde T_{div}(M^2)\int_k \frac{i}{(k^2-M^2)^2}\nonumber\\  
&=&  
\frac{i}{\gamma}\left[T_d^{(2)}+(4M^2-M^2_0)T_d^{(0)}\right]-i\left[\frac{1}{\gamma}-I^F(0,M)+\frac{1}{8\pi^2}\right]\left(\tilde{T}_{div}-3\gamma M^2T_a^{(I)}\right) \nonumber\\  
&& -i\tilde{T}_{div}(M^2)\left(T_d^{(0)}+I^F(0,M)\right).  
\end{eqnarray}  
The terms proportional to $I^F(0,M)$ can be rewritten with the help of a relation between the  
finite part of the tadpole integral $T^F(M)$ and the bubble integral at  
zero external momentum:  
\begin{equation}  
M^2I^F(0,M)=T^F(M)+\frac{1}{16\pi^2}(M^2-M^2_0).  
\end{equation}  
In this way one finds a piece linear in $M^2$ with somewhat  
complicated looking divergent coefficients and a dangerous term resulting from an uncancelled subdivergence:  
\be  
I_3=i(A_{div}+B_{div}M^2-3\gamma T^F(M) T_a^{(I)}),  
\label{Eq:div-i3}  
\ee 
where
\begin{equation}
A_{div}=\frac{1}{\gamma}\Big[T_d^{(2)}-M_0^2T_d^{(0)}-\tilde{T}_{div}\Big]+\frac{3}{16\pi^2}\gamma T_a^{(I)} M_0^2-\Big[\frac{1}{8\pi^2}+T_d^{(0)}\Big]\tilde{T}_{div},
\nonumber\\
\end{equation}
\begin{equation}
B_{div}=\frac{4}{\gamma}T_d^{(0)}+\Big[3+\frac{3\gamma}{16\pi^2}\Big]T_a^{(I)}.
\end{equation}

\renewcommand{\theequation}{B\arabic{equation}}

\section{Useful relations involving the $U(n)$ structure constants}
In the following relations, the summation over index $e$ (where it appears) goes from $0$ to $n^2-1$.
\begin{equation} \nonumber\\
f_{0ab}=0, \qquad d_{0ab}=\sqrt{\frac{2}{n}}\delta_{ab}, \qquad \sum_{i=1}^{n^2-1} d_{0ii}=(n^2-1)\sqrt{\frac{2}{n}}, \qquad \sum_{i=1}^{n^2-1} d_{iia}=\delta_{0a}(n^2-1)\sqrt{\frac{2}{n}},
\end{equation}
\begin{equation}
\sum_{i=1,j\neq 0}^{n^2-1}d_{iij}=0, \qquad
\sum_{a=0}^{n^2-1}d_{aaf}=\sqrt{2n^3}\delta_{f0}, \qquad
\sum_{a=0}^{n^2-1}d_{aae}d_{ecd}=2n\delta_{cd}, \qquad
\sum_{i=1}^{n^2-1}d_{iie}d_{ecd}=\frac{2}{n}(n^2-1)\delta_{cd}, \nonumber\\
\end{equation}
\begin{equation}
\sum_{i,j=1}^{n^2-1}d_{ijc}d_{ijd}=n\Big[1+\delta_{c0}\delta_{d0}-\frac{2}{n^2}(2-\delta_{c0}\delta_{d0})\Big]\delta_{cd}, \qquad
\sum_{i=1,a=0}^{n^2-1}d_{aic}d_{aid}=n\Big[1-\frac{2}{n^2}+\delta_{c0}\delta_{d0}\Big]\delta_{cd},
\nonumber\\
\end{equation}
\begin{equation}
\sum_{a=0}^{n^2-1}d_{ace}d_{ade}=n(1+\delta_{c0}\delta_{d0})\delta_{cd}, \qquad
\sum_{i=1}^{n^2-1}f_{ice}f_{ide}=n(1-\delta_{c0}\delta_{d0})\delta_{cd}.
\end{equation}

 \end{document}